# Theoretical prediction of the importance of the $^3B_2$ state in the dynamics of sulfur dioxide.

*Camille Lévêque* * [1,2,3], *Richard Taïeb*[1,2] *and Horst Köppel*[3]

[1] Sorbonne Universités, UPMC Univ. Paris 06,

Laboratoire de Chimie Physique-Matière et Rayonnement, UMR 7614,

11 Rue Pierre et Marie Curie, 75231 Paris Cedex 05, France.

[2] CNRS, LCPMR, UMR 7614, Paris Cedex 05, France

[3] Theoretische Chemie, Physikalisch-Chemisches Institut, Universität Heidelberg, Im Neuenheimer Feld 229, D-69120 Heidelberg, Germany

**Corresponding Author**

**Camille Lévêque**

Camille.Leveque@pci.uni-heidelberg.de

While, the sulfur dioxide molecule has been extensively studied over the last decades, its photoexcitation dynamics is still unclear, due to its complexity, combining conical intersections and spin-orbit coupling between a manifold of states. We present for the first time a comprehensive ab initio study of the intersystem crossing of the molecule in the low energy domain, based on a wave-packet propagation on the manifold of the lowest singlet and triplet states. Furthermore, spin-orbit couplings are evaluated on a geometry-dependent grid, and diabatized along with the different conical intersections. Our results show for the first time the primordial role of the triplet $^3B_2$ state which has not been discussed before, giving new insights into the dynamics of the intersystem crossing in $SO_2$.

The sulfur dioxide molecule has been widely studied over the last century [1-6], due to its preponderant role in the photophysics and photochemistry of the earth's atmosphere [7-11], and other planets [12,13]. Experimentally [14,15], it is known for a long time that only the triplet states can give rise to chemical reactions with other atmospheric compounds, the lowest singlet states being just intermediates, thanks to the dipole-allowed transition from the ground state to one of them ($^1B_1$). The aerosol products raised an enormous interest in the S-MIF effect [16-20], to provide a way to understand the Archean sediment isotopic composition. The three lowest triplet states are known to react with different molecules such as CO, NO, $O_2$, $C_2H_2$, allene, biacetyl... Note that only indirect observations of the $^3A_2$ and $^3B_2$ states were performed while the $^3B_1$ state was directly observed in phosphorescence and photoabsorption spectra [21].

To this day only two attempts to understand the intersystem crossing have been undertaken in the literature. The first one [22], based on surface-hopping techniques and a constant SOC to describe the photodynamics of the molecule on the manifold of three lowest triplet states ($^3B_1$, $^3A_2$ and $^3B_2$) and the two lowest singlet excited states ($^1B_1$ and $^1A_2$), shows a very fast (only 30 fs) and efficient (40%) transfer to the triplet states. The second description [23] uses quantum dynamics for the intersystem crossing between the former singlet states and *only* two triplet states ($^3B_1$ and $^3A_2$) and gives rise to similar transfer, but in 1ps with only ~10% reached in 200fs. In this study, the authors committed two mistakes and two crude approximations (see [24]) by first discarding the $^3B_2$ state and by wrongly introducing spin-orbit coupling (SOC) between the $^3B_1$,and $^1B_1$ states, and further by ignoring the conical intersection between the remaining triplet states and considering constant SOC.

In this letter, we provide a reliable comprehension of the intersystem crossing in $SO_2$ and clarify the contribution of the different electronic states, where it turns out that the $^3B_2$ state plays a key role. Furthermore, we evidence the role of interfering SOC channels in the population of this triplet state.

To describe accurately the system, the potential energy surfaces (PES) involved in the dynamics of $SO_2$ have been computed using the multi-reference configuration interaction (MRCI) method. In addition, we computed for the first time a three dimensional map of the spin-orbit coupling, using the Breit-Pauli operator and MRCI electronic wave-functions, to account for the geometry dependence of this latter. All ab initio calculations have been performed using the MOLPRO program suite [25]. When considering the degeneracy removal occurring within the triplet manifold, the system considered is then described by 12 states. Using symmetric (+) and asymmetric (-) combinations of spin basis functions with $M_S = \pm 1$, we show that these states belong to one of the two irreducible representations, i.e. A' and A", of the $C_s$ point group. As the ground electronic state of the molecule has A' symmetry, the one photon excitation will populate excited singlet states transforming as A". Due to the block-diagonal structure of the twelve by twelve Hamiltonian matrix,

$$H = T.\mathbf{1}_{12} + \begin{pmatrix} V_{A'} & 0 \\ 0 & V_{A''} \end{pmatrix},$$

no interaction occurs between the A' and A" states after the initial stage of the photoexcitation. Here, $T$ is the nuclear kinetic energy operator, $\mathbf{1}_{12}$ the 12 × 12 unit matrix and $V_{A', A''}$ are the potential energy matrices including the SOCs for each symmetry. This block structure allows us to reduce the number of states in our study by considering only the block A", which includes the following electronic states $\{^1B_1, ^1A_2, ^3B_1(-), ^3A_2(-), ^3B_2(+), ^3B_2(0)\}$. The main challenge is to treat correctly the two symmetry-allowed conical intersections (CI) occurring between the triplet states ($^3B_1, ^3A_2$) and the singlet states ($^1B_1, ^1A_2$), respectively. The latter was addressed in our previous study [26] and we use here the same method for both, i.e. we make use of the regularized diabatization scheme introduced in [27]. Collecting the nuclear coordinates in $Q = (R_S, \phi, Q_u)$, with $R_s = (R_1 + R_2)/2$, $Q_u = (R_1 - R_2)/2$ and $\phi$ with $R_{1;2}$ and $\phi$ being the two SO bond lengths and the OSO angle, respectively, the diabatic potential matrix reads

$$W_{reg}^i(Q) = \Sigma^i(Q)\,1_2 + \frac{\Delta^i(Q)}{\sqrt{(\Delta_0^i)^2(R_S,\phi) + W_{12}^2}} \cdot \begin{pmatrix} \Delta_0^i(R_S,\phi) & W_{12}^i(Q_u) \\ W_{12}^i(Q_u) & -\Delta_0^i(R_S,\phi) \end{pmatrix}.$$

The superscript $i$ refers to the two CIs, i.e. within the singlet and the triplet manifold.

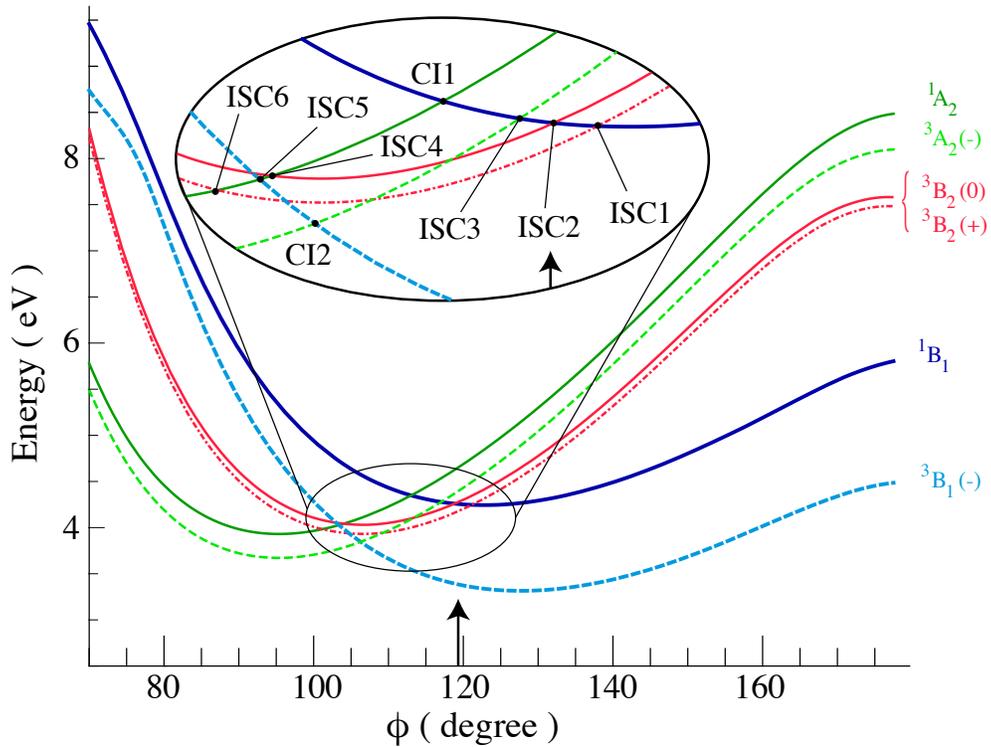

**Figure 1.** One-dimensional cut through the PES, of the A'' states, along the bending angle $\phi$. The two singlet states, $^1B_1$ ($^1A_2$), are represented in thick dark blue (thin green), respectively. In the same way, the triplet $^3B_1$ ($^3A_2$) states are represented in thick dashed light blue (thin dashed light green). The triplet $^3B_2$ state has two different spin components, (0) or (+), both represented in full and dot-dashed red, respectively, with an artificial shift to separate them. In the inset we magnify the central region where ICS and CI occur.

We introduced $\Sigma^i(Q) = (V_1^i(Q) + V_2^i(Q))/2$, $\Delta^i(Q) = (V_1^i(Q) - V_2^i(Q))/2$ and $\Delta_0^i(R_S, \phi) = \Delta^i(R_S, \phi, Q_u = 0)$, the half-sum and half-difference of the adiabatic potentials, and, the value of $\Delta^i(Q)$ in the $C_{2v}$ subspace, respectively. In addition, $W_{12}^i(Q_u) = \lambda^i Q_u$, in which $\lambda^i$ is defined as

$$\lambda^i = \frac{\partial((V_1^i(Q) - V_2^i(Q)))}{\partial Q_u}\bigg|_{Q_{seam}},$$

and has been evaluated along the seam of conical intersections. The SOC have been computed between the different adiabatic electronic states and have to be transformed into the diabatic basis. This issue arises from the geometry-dependence of the SOC mainly due to the drastic change of the electronic wave function from one side to the other of the two CI and the change of symmetry from $C_{2v}$ to Cs, during the dynamics, making some forbidden couplings allowed. To diabatize the SOC elements we used the diabatic-to-adiabatic transformation matrix, $S(Q)$, which reads for the A" system:

$$S(Q) = \begin{pmatrix} S_S & 0 & 0 \\ 0 & S_T & 0 \\ 0 & 0 & 1_2 \end{pmatrix},$$

where $S_S$ and $S_T$ are the two-by-two transformation matrices for the singlet and triplet states respectively [26, 27]. As the $^3B_2$ states are not <u>vibronically</u> coupled with the others, the $2 \times 2$ unit matrix ($1_2$) is introduced in the full transformation matrix, $S(Q)$. Fore more details see Ref. [28].

The wavepacket propagation is then carried out using the MCTDH package [29] to integrate the time-dependent Schrödinger equation within the 6 coupled-states A" manifold, see Fig. 1, the initial wave-function being the molecular vibrational ground state vertically launched onto the diabatic $1^1B_1$ state, the only dipole allowed transition. Note that we use a direct expansion of the nuclear wave packet (wp) on a DVR basis, instead of the MCTDH wave function.

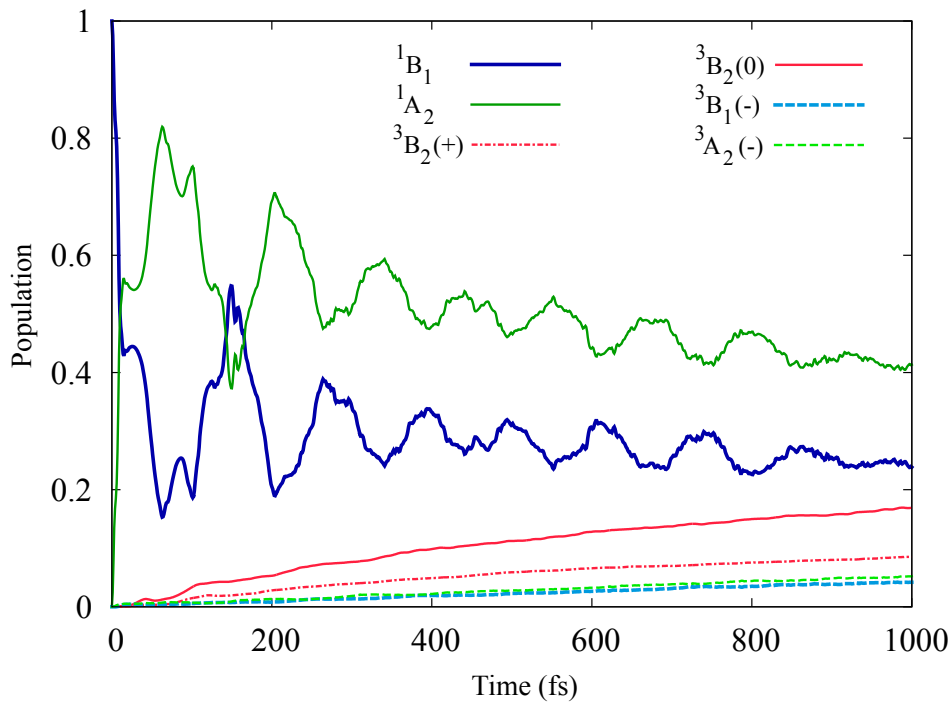

**Figure 2.** (Color online) Populations on the different electronic states as a function of time. The same color code as in fig. 1 is used. In thick dark blue and thin dark green are the population in the electronic singlet $^1B_1$ and $^1A_2$, respectively. The major part of the population into the triplet manifolds belongs to the $^3B_2(0)$, in full red and $^3B_2(+)$, in dot-dashed red. The remaining density is shared between the $^3B_1(-)$, in thick dashed light blue and $^3A_2(-)$ in thin dashed light green.

We report in Fig. 2, the time-dependent electronic populations of the full wave-function integrated over the different electronic states. The singlet state population exhibits oscillations with a period of ~150fs, reminiscent of what we obtained in [26]. We point out that, after 1ps, the population is quite different to what was observed in previous studies [22, 23], with approximately 24% in $^1B_1$, 41% in $^1A_2$, 25% in the $^3B_2$ states (17% in the $^3B_2(0)$ and 8% in the $^3B_2(+)$), and the remaining triplet states summing to less than 10% (4.2% in $^3B_1(-)$ and 5.3% in $^3A_2(-)$).However, because only adiabatic triplet populations are given in [22], a detailed comparison on the basis of individual triplet states is not possible and, in particular, symmetry assignments, inherently diabatic in nature, cannot easily be extracted from the results of Ref. [22]. While a fast intersystem crossing is observed, within the first 100 fs, followed by a slow transfer from the singlet to the triplet states, it does take place on a longer timescale than the non-adiabatic coupling (~20 fs), in contrast to the conclusions of [22].

To understand the underlying mechanism for the singlet-triplet conversion, we concentrate hereafter on the first 200 fs of the propagation, a time for which the wp does not spread much over the singlet state surfaces but remains localized [26]. In the following, we will refer to the different crossings or CIs that are labeled in Fig. 1. We report in Fig. 3 the population of the $^3A_2$ and $^3B_1$ states as well as their sum. The latter is a relevant quantity because the conical intersection (CI2) between these two states allows the wp to move from one state to another and could hide the dynamics of the SOC-triggered transfer. We have also added the autocorrelation function $C(t) = |\langle\Psi(0)|\Psi(t)\rangle|$, with $\Psi(t)$, being the full wavefunction of the system. We clearly see three steps during the internal conversion occurring for times between 0 and 16 fs, 55 and 65 fs and 150 and 165 fs, respectively. The first step is attributed to a $^1B_1 \rightarrow{}^3A_2$ transfer, when the nuclear wp, launched on the $^1B_1$ state follows the $^1B_1/^3A_2$ intersection (ISC3). Then the non-adiabatic coupling between $^3A_2$ and $^3B_1$ (CI2), will transfer in a few fs (~7fs) the population from the former to the latter, but the sum of populations remains constant until t~55 fs. During this time the majority of the singlet population is transferred to $^1A_2$ through CI1 [26], see also Fig. 2. Thus, the second step comes from the $^1A_2 \rightarrow{}^3B_1$ interaction, when the wp comes back to near the Frank-Condon (FC) area, but with a smaller angle (110°) [26], which is located in the vicinity of the $^1A_2$-$^3B_1$ intersection (ISC5). This dynamics of the wp is confirmed by the revival in *C(t)* at *t*~65 fs. More interestingly, the third step takes place during the 16 fs following the next revival in *C(t)* at *t*~150 fs, when the wp returns to a geometry very close to the initial one (*t*=0) [26], allowing the same transfer as step 1. The stepwise shape of the dynamics of the singlet/triplet transfer demonstrates the primordial role of the crossing of the PES, due to the smallness of the SOC elements, which allows intersystem crossing only at localized geometries on the PES.

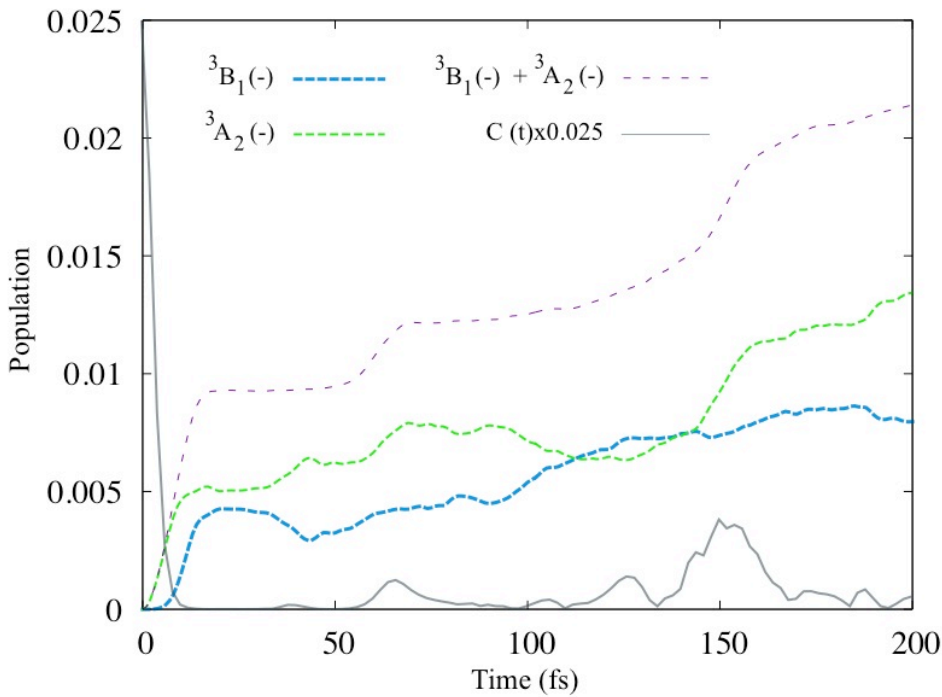

**Figure 3.** The population on $^3B_1(-)$ (thick dashed blue) and $^3A_2(-)$ (thin dashed green) electronic states is represented for the first 200fs. In addition their sum is represented in dashed purple. In light grey, we show the auto-correlation function.

The above explanation is supported by simulations where we artificially switched off the SOC between either $^1A_2/^3B_1$ or $^1B_1/^3A_2$. In the first case, only the first and third transfers are observed, whereas in the second case only the transfer at 55<t<65 fs occurs. Finally, we would like to emphasize that the $^3B_2$ states are not involved, during this time, in the mechanism populating the $^3B_1/^3A_2$ states.

We follow the same reasoning to understand the population transfer to the two $^3B_2$ states of A" symmetry, which present a similar slightly smoothed step-like time evolution shown in Figs. 4. Considering first the $^3B_2(+)$, three transfers take place in the first 80 fs, indicated in Fig.

4. The first one (small) occurs between 0 and 7 fs, is only due to the coupling between the $^1B_1 \to ^3B_2(+)$ (ISC1). The two others are governed almost exclusively by the interaction between $^1A_2$ and $^3B_2(+)$ (ISC6). Their efficiency is explained by, apart from a slightly stronger SOC, the dynamics of the wp on the $^1A_2/^1B_1$ states. The wp between 20 to 40 fs is at the turning point of the asymmetric mode ($Q_U$) [26] spending then a longer time near the crossing between $^1A_2$ and $^3B_2$ (ISC4,6). This happens again when the wp moves back to the FC area before the revival at 150 fs. The time spent in the direct vicinity of the ISC6 seam also gives a back-transfer to the singlet states, at $t$~55 fs. This was not the case for the other triplet states ($^3A_2/^3B_1$) due to the location of the ISC3 and 5. These different geometry configurations also explain the washing-out of the step as a part of the wp is continuously at or near the ISC6 and 4 seam, when considering the $^3B_2$ states.

Similar behavior explains the population transfer to the $^3B_2(0)$ state which features the same PES as $^3B_2(+)$. While the difference of the populations between the two $^3B_2$ states could come from different SOC with the $^1A_2$ state (~45 cm$^{-1}$ for $^3B_2(0)$ and ~35 cm$^{-1}$ for $^3B_2(+)$), this is not enough to explain a ratio of ~1.9 at 300fs. Such enhancement arises from interfering channels populating the $^3B_2(0)$ state. To demonstrate this effect, we performed simulations when we switched off either the $^1A_2 \to ^3B_2(0)$ or the $^1B_1 \to ^3B_2(0)$ SOC or changing the sign of the latter. We show the results in Fig. 4b, where one sees that the incoherent sum of the two channels does not reproduce the full calculation while the change of sign exhibits strong destructive interferences compared to the coherent sum, which almost matches the full calculation. Again to discard the role of the $^3A_2$ and $^3B_1$ for t<200fs, we removed them from our calculations and no impact on our results is visible, except for longer times 200fs<t<1ps.

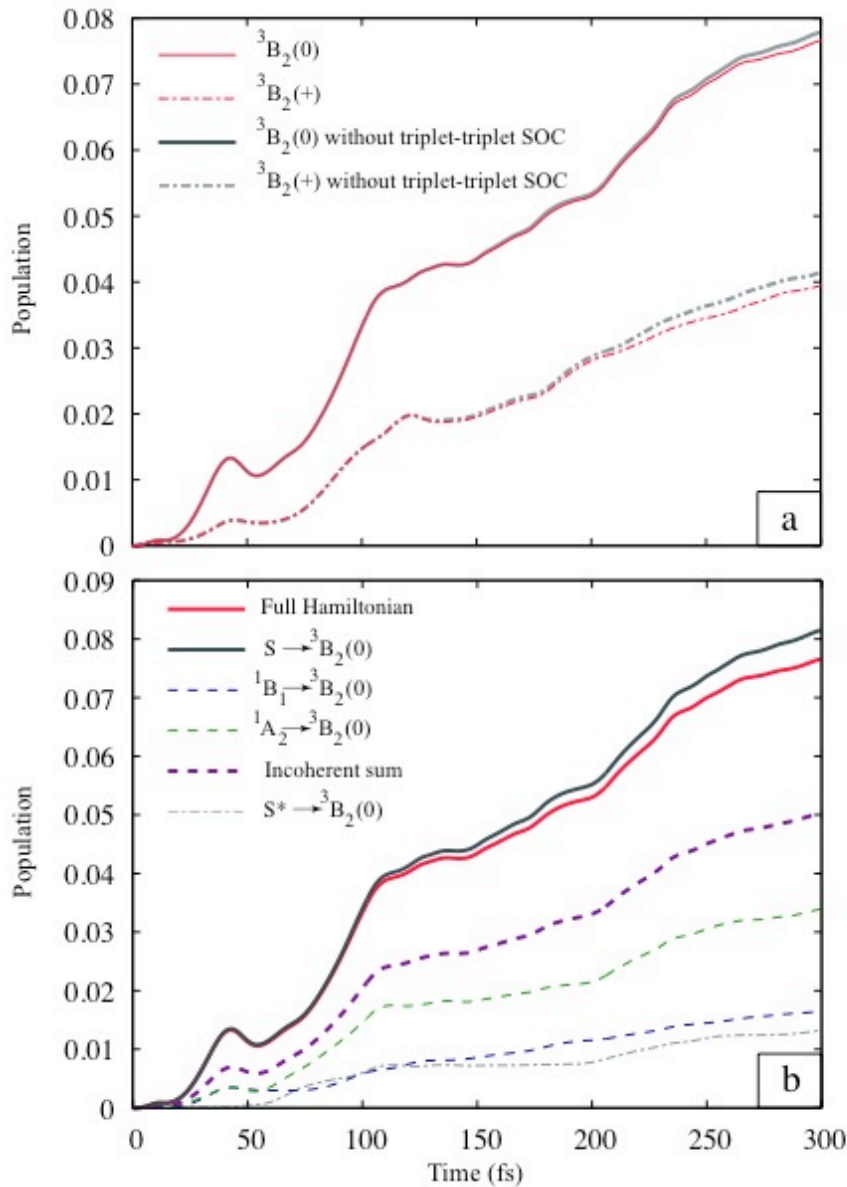

**Figure 4.** a) The population on $^3B_2(+)$ (thin dot-dashed red) and $^3B_2(0)$ (thin full red) electronic states is represented for the first 300fs. The thick (dot-dashed) grey curves present the analogous results obtained if we neglect the SOC between all the triplet states. b) In full thick red is represented the result from the full Hamiltonian in addition the full thick grey line is the population when only the coupling between the singlet ($^1A_2/^1B_1$) states and $^3B_2(0)$ is considered. In thin dashed blue (thin dashed green) we represent the population when only SOC between $^1B_1$ ($^1A_2$) and $^3B_2(0)$ is used. The incoherent sum of these former populations is represented in thick dashed purple. The destructive interference (dot-dashed thin grey) is obtained by changing arbitrarily the sign of the coupling between $^1B_1$ and $^3B_2(0)$.

In this work, the role of the triplet states in the photophysics of $SO_2$ has been investigated. We clearly demonstrated, for the first time, that the $^3B_2(0)$ and $^3B_2(+)$ states play a major role in the intersystem crossing. Our findings, based on a full quantum description of the system, are strongly different from previous ones. Unfortunately no direct comparison with experiment is possible. This state has been already theoretically predicted by I. Hillier *et al*. [5], but has never been observed experimentally. Corresponding experiments are strongly encouraged to confirm our predictions.

## Acknowledgments

Hans-Jakob Wörner, David Villeneuve, Albert Stolow and Leticia González are warmly thanked for stimulating discussions. Financial support by the Max-Planck-Institute für Kernphysik Heidelberg through the IMPRS Quantum Dynamics in Physics, Chemistry and Biology is gratefully acknowledged. Support from the ANR-09-BLAN-0031-01ATTO-WAVE is also acknowledged. A part of the computations has been performed at the bwGRiD (http://www.bw-grid.de), member of the German D-Grid initiative, funded by the Ministry for Education and Research (Bundesministerium für Bildung und Forschung) and the Ministry for Science, Research and Arts Baden-Württemberg (Ministerium für Wissenschaft, Forschung und Kunst Baden-Württemberg).